\documentclass{ws-p9-75x6-50}

\usepackage{amsmath,psfrag}

\begin{document}

\title{Dynamics of a Large--Spin--Boson System in the Strong
       Coupling Regime}

\author{Till Vorrath}
\address{I. Inst. f. Theor. Physik, Universit\"at Hamburg,
Jungiusstr. 9, 20355 Hamburg, Germany \\E-mail: vorrath@physnet.uni-hamburg.de}
\author{Tobias Brandes}
\address{Dept. of Physics, UMIST, P.O. Box 88, Manchester M60 1QD, UK}
\author{Bernhard Kramer}
\address{I. Inst. f. Theor. Physik, Universit\"at Hamburg,
Jungiusstr. 9, 20355 Hamburg, Germany}


\maketitle

\abstracts{We investigate collective effects of an ensemble of biased two
level systems interacting with a bosonic bath in the strong coupling regime.
The two level systems are described by a large pseudo-spin $J$.
An equation for the expectation value $M(t)$ of the $z$-component of the 
pseudo spin is derived and solved numerically for an ohmic bath at $T\!=\!0$.
In case of a large cut-off frequency of the spectral function, a Markov
approximation is justified and an analytical solution is presented.
We find that $M(t)$ relaxes towards a highly correlated 
state with maximum value $\pm J$ for large times. However, this relaxation is 
extremely slow for most parameter values so as if the system was "frozen in"
by interaction with the bosonic bath.}
\section{Introduction}
The spin-boson model deals with a biased two level system, described as a 
spin $\frac{1}{2}$, under the influence of a bosonic environment 
\cite{leggett}. It is applied to different physical systems
such as, e. g.,  two level systems in amorphous materials, or
double quantum dots interacting with phonons \cite{kouwenhoven,brandes,icps}
or photons. 
Further examples include a magnetic flux in a SQUID which originally 
motivated the spin-boson model \cite{caldeira}.

In this contribution, we study the more general case of many identical
two level systems (or larger spins) that couple to a bosonic field.
Examples are glasses or arrays of coupled dots.
We focus on collective effects and hence, investigate the dynamics of 
the total spin with projection $J_z=\sum_i \sigma_i^z$.
This pseudo spin gives information about the degree of polarization, 
i. e. the number of two level systems to be found in each of the two states.
We concentrate on the strong coupling regime, when perturbation
theory is not sufficient to take into account the interaction with the
environment.
\section{The large--spin--boson model}
The large spin-boson system is described by the Hamiltonian
\begin{equation}
H = \varepsilon J_z + 2T_c J_x + J_z \sum_q \gamma_q (a_{-q} + a_q^+)
    + \sum_q \omega_q \, a_q^+ a_q,
\end{equation}
which is formally equivalent to the spin-boson Hamiltonian. The only 
difference is that $J$, obeying the spin algebra 
$[J_i,J_j]=\varepsilon_{ijk} i  J_k$,
is not restricted to a spin $\frac{1}{2}$ but can take any value.
$\varepsilon$ and $T_c$ are the energy difference and the tunnel rate, 
respectively, between the two states of the two level systems.
The environment is given by the bosonic field operators $a_q$,
eigenenergies $\omega_q$ ($\hbar\!=\!1$) 
and interaction coefficients $\gamma_q$ for the mode $q$.
To take into account arbitrarily strong couplings between the large spin
and the bosonic bath, a polaron transformation is applied to the Hamiltonian,
\begin{equation}
\label{ham-polaron}
\bar{H} = e^{\sigma J_z} H e^{- \sigma J_z}
        = \varepsilon J_z -\alpha J_z^2 + \sum_q \omega_q \, a_q^+ a_q
        + T_c (J_+ \,X + J_- \,X^+),
\end{equation}
\begin{equation}
\hspace*{-29.5mm}\mbox{with} \qquad
\sigma := \sum_q \frac{\gamma_q}{\omega_q} \, (-a_{-q} + a_q^+), \quad 
     X := e^{\sigma}, \quad
\alpha := \sum_q \frac{|\gamma_q|^2}{\omega_q}.
\end{equation}
As a result of the polaron transformation the interaction term between the
spin and the environment disappears whereas the tunnel term $2T_c J_x$ becomes
dependent on the bosonic field. Furthermore, a new term $-\alpha J_z^2$
appears in the transformed Hamiltonian which will later turn out to be of 
importance for the long time behaviour of the large spin. For a single 
two level system, i.e. $J=\frac{1}{2}$, this term does not play any role 
since $J_z^2$ is constant in that case.

Next, the transformed Hamiltonian (\ref{ham-polaron}) is used 
to derive a master equation for the density matrix $\rho(t)$ of the spin.
In second order Born approximation \cite{carmichael}, its time evolution
is given by
\begin{equation}
\label{master}
\dot{\tilde{\rho}}(t) = - \int_0^t \! dt' \; Tr_{Res} \!
   \left\{[\tilde{H}_T(t),[\tilde{H}_T(t'),\tilde{\rho}(t')\otimes R_0]]
     \right\},
\end{equation}
where $\tilde{H}_T(t)$ is the the tunnel part of the Hamiltonian 
in interaction picture. 
In this equation, only a product of the spin density matrix $\rho$
and the density matrix of the bosonic bath $R_0$ (which is assumed to be in
thermal equilibrium at all times) appears. 
This makes it possible to trace over the
reservoir's degrees of freedom without knowing the full density matrix 
of the combined system of spin and environment
at all times. Note that $\tilde{\rho}(t')$ is evaluated at times 
$t'$ in the integral in Eq. (\ref{master}), thus describing a 
non-Markovian process.
Although we used the second order Born approximation, the interaction between
spin and bath is taken into consideration to arbitrary high orders due to
the polaron transformation.
A closed set of equations for the diagonal elements $\rho_{M,M}$ in the 
basis $\left|JM\right\rangle$ of eigenstates of $J^2$ and $J_z$ is derived 
from the master equation (\ref{master}) under the assumption that the 
elements $\rho_{M,M\pm2}$ can be neglected.
These correspond to simultaneous spin flips of two spins and hence two
boson processes.
In the semiclassical limit of many two level systems and therewith
large spin $J$ \cite{gross}, 
the variance of $J_z$ can be neglected and we find for the expectation value 
$M(t)\equiv \left\langle J_z \right\rangle_t = \sum_M M \rho_{M,M}(t)$,
\begin{equation}
\label{m}
\begin{split}
\dot{M}(t) = 
 -2 T_c^2 \int_0^t \!\! &dt' \big(J\!+\!M(t')\big)\big(J\!-\!M(t')\!+\!1\big)\;
 \mbox{Re}\left\{e^{i(\varepsilon - \alpha (2M(t')-1))(t\!-\!t')} 
    C(t\!-\!t')\right\}\\
    &-\big(J\!+\!M(t')\!+\!1\big)\big(J\!-\!M(t')\big)\;
 \mbox{Re}\left\{e^{-i(\varepsilon - \alpha (2M(t')+1))(t\!-\!t')} 
   C(t\!-\!t')\right\}\!.
\end{split}
\end{equation}
\section{The bosonic bath}
The correlation function
\begin{equation}
C(t\!-\!t') = Tr_{Res} \left\{ R_0 \; \tilde{X}_t \tilde{X}_{t'}^+ \right\}
\end{equation}
describes
the influence of the environment on the large spin in Eq. (\ref{m}).
It acts as a memory that returns information about the spin at time $t'$
back to the spin at time $t$ after the delay $t\!-\!t'$.
As usual, we assume that the spectral function $J(\omega)$, 
\begin{equation}
J(\omega) = \sum_q |\gamma_q|^2 \; \delta(\omega-\omega_q)\, ,
\end{equation}
exhibits a power law behaviour $J(\omega) = g \; \omega^s e^{-\omega/\omega_c}$
with an exponential cut-off at frequency $\omega_c$, 
and $g$ the effective interaction strength. Then, the correlation function
can be calculated analytically. In the case of an ohmic bath ($s=1$)
and at temperature $T=0$, the correlation function reads 
\begin{equation}
C(t) = \left( 1 + i \omega_c t \right)^{-g} \, .
\end{equation}
For the limit of an infinite cut-off frequency $\omega_c$, we find
\begin{equation}
\lim_{\omega_c \to \infty} \omega_c \; \mbox{Re} 
     \big\{ e^{i\varphi \omega_c t} C(t)  \big\}
   = \frac{2 \pi \varphi^{g-1}}{\Gamma(g) e^{\varphi}} \;\Theta(\varphi) \;
     \delta(t).
\end{equation}
Thus, the Markov approximation is justified for large cut-off frequencies 
where analytic solution of Eq. (\ref{m}) becomes possible. For $g\!=\!1$ the 
inverse function $t=t(M)$ is
\begin{equation}
\label{analytisch}
t(M) = \begin{cases}
       ((2J\!+\!1)\Gamma)^{-1}\,(f(M)-f(M_0)), 
               &M_0 < -\frac{1}{2}+\frac{\varepsilon}{2\omega_c} \\[1mm]
       ((2J\!+\!1)\tilde{\Gamma})^{-1}\,(f(-M)-f(-M_0)), 
               &M_0 > \frac{1}{2}+\frac{\varepsilon}{2\omega_c}
       \end{cases}
\end{equation}
with $\Gamma = e^{-\varepsilon/\omega_c-1} 2 \pi T_c^2 / \omega_c $,
$\tilde{\Gamma} = e^{\varepsilon/\omega_c-1} 2 \pi T_c^2 / \omega_c $
and 
\begin{equation}
f(x)= e^{-2(J+1)}\,\mbox{Ei}(-2x\!+\!2(J\!+\!1))-e^{2J}\,\mbox{Ei}(-2x\!-\!2J).
\end{equation}
Ei$(x)$ is the exponential-integral function and $\alpha\!=\!g\omega_c$.
Initial values $M_0$ in the interval 
$M_0\in[-1/2+\varepsilon/(2\omega_c), 1/2+\varepsilon/(2\omega_c)]$ 
are not considered here, as this interval is of width 1 and 
hence contains only one possible value of $M_0$.
\section{Results}
\begin{figure}
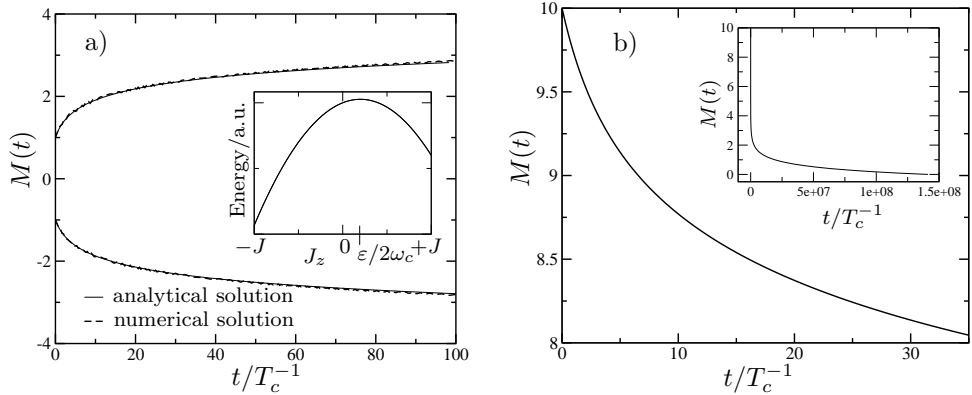

\psfrag{a}{a)}
\psfrag{t}{\hspace*{-3mm}$t/T_c^{-1}$}
\psfrag{M(t)}{$M(t)$}
\psfrag{Ana}{\footnotesize analytical solution}
\psfrag{Num}{\footnotesize numerical solution}
\psfrag{s}{\footnotesize $\varepsilon/2\omega_c$}
\psfrag{-J}{\footnotesize \hspace*{-3mm} $-J$}
\psfrag{+J}{\footnotesize \hspace*{-3mm} $+J$}
\psfrag{N}{\footnotesize $0$}
\psfrag{z}{\footnotesize $J_z$}
\psfrag{E}{\footnotesize \hspace*{-8mm} Energy/a.u.}
\epsfig{file=fig1.eps,width=6cm}
\hfill
\psfrag{b}{b)}
\psfrag{t1}{\hspace*{-5mm}$t/T_c^{-1}$}
\psfrag{M(t)}{$M(t)$}
\psfrag{t2}{\hspace*{-1.5mm}\footnotesize$t/T_c^{-1}$}
\psfrag{M2}{\footnotesize \hspace*{-2mm}$M(t)$}
\epsfig{file=fig2.eps,width=6cm}
\caption{a shows $M(t)$ for eight two level systems, i. e. $J\!=\!4$, with
$\varepsilon\!=\!T_c$, $\omega_c\!=\!20T_c$ and $g\!=\!1$ for initial values
$M_0\!=\!\pm1$. The analytical solution (Eq.\ref{analytisch}) and the 
numerical solution of Eq.(\ref{m}) coincide, indicating that the Markov 
approximation is justified. The inset shows the spin system eigenenergies
as explained in the text.
b: Analytical solution for $M(t)$ for  $J\!=\!10$,
$\varepsilon\!=\!440T_c$, $\omega_c\!=\!20T_c$ and $g\!=\!1$.
After an initial phase, the decay is strongly damped as can be seen from
the inset, which shows $M(t)$ for larger times.}
\label{bild}
\end{figure}
We solved the integro-differential equation (\ref{m}) for the expectation
value $M(t)$ numerically. This enables us to check the Markov approximation
which was applied to derive the analytical solution (\ref{analytisch}).
In case the assumptions for the derivation of Eq. (\ref{analytisch})
are not fulfilled ($g\ne 1$ or small $\omega_c$), results can only be
obtained by the numerical solution.
Fig. \ref{bild}.a illustrates that the inversion $M(t)$ approaches $+J$
or $-J$ for large times depending on the initial value of $M_0$. 
If $M_0 >\varepsilon/2\omega_c$, we find that for large times $t$, 
$M(t)$ relaxes towards a state with macroscopic complete inversion $+J$
which means that all spins are up in spite of $\varepsilon > 0$.
This behaviour is due to the spin part $\varepsilon J_z-\alpha J_z^2$
of the transformed Hamiltonian (\ref{ham-polaron}), the eigenvalues of 
which lie on an inverted parabola with two minima at the extreme points
$\pm J$ (inset of Fig. \ref{bild}a).
Therefore, for very strong coupling to the bosonic field, the total system
minimizes its energy by relaxing towards either to the global ($M\!=\!-J$) 
or local ($M\!=\!+J$) minimum.
This situation is analogous to the emergence of a macroscopically
polarized state in the strong coupling limit of the one--mode 
superradiance model \cite{hepp,wang}.
However, the dynamics of this relaxation is extremely slow if $M(t)$
is not close to $\varepsilon / 2 \omega_c$ as can be seen
from Fig. \ref{bild}.b (the parameters were chosen such that $M(t)$
decays to $-J$ even for the initial value $M_0\!=\!+J$).
Thus, the system seems to be "frozen in" by the interaction 
with the bosonic bath for most parameter values.


\begin{thebibliography}{99}
\bibitem{leggett} A. Leggett \etal, \Journal{\RMP}{59}{1}{1987}.
\bibitem{kouwenhoven}  T. Fujisawa \etal, \Journal{\Sci}{282}{932}{1998}.
\bibitem{brandes} T. Brandes and B. Kramer, \Journal{\PRL}{83}{3021}{1999}.
\bibitem{icps} S. Debald, T. Vorrath, T. Brandes, B. Kramer, 
    \emph{Proc. 25th Int. Conf. Phys. Semicond., Osaka 2000}, 1049  
    (Springer-Verlag, Berlin, 2001).
\bibitem{caldeira} A. Caldeira and A. Leggett, \Journal{\PRL}{46}{211}{1981}.
\bibitem{carmichael} H. Carmichael, \emph{An Open Systems Approach 
    to Quantum Optics} (Springer-Verlag, Berlin, 1993).
\bibitem{gross} M. Gross and S. Haroche, \Journal{\PRep}{93}{302}{1982}.
\bibitem{hepp} K. Hepp and E. Lieb, \Journal{\AP}{76}{360}{1973}.
\bibitem{wang} Y. K. Wang and F. T. Hioe, \Journal{\PRA}{7}{831}{1973}.
\end{thebibliography}
\end{document}